# Supporting Acceptance Testing in Distributed Software Projects with Integrated Feedback Systems: Experiences and Requirements

Olga Liskin*, Christoph Herrmann†, Eric Knauss‡, Thomas Kurpick†, Bernhard Rumpe†, and Kurt Schneider*
* *Software Engineering Group, Leibniz Universität Hannover, Germany*
*Email: {olga.liskin,kurt.schneider}@inf.uni-hannover.de*
† *Software Engineering, RWTH Aachen University, Germany*
*Web: http://www.se-rwth.de*
‡ *University of Victoria, Canada, Email: knauss@computer.org*

*Abstract*—During acceptance testing customers assess whether a system meets their expectations and often identify issues that should be improved. These findings have to be communicated to the developers – a task we observed to be error prone, especially in distributed teams. Here, it is normally not possible to have developer representatives from every site attend the test. Developers who were not present might misunderstand insufficiently documented findings. This hinders fixing the issues and endangers customer satisfaction. Integrated feedback systems promise to mitigate this problem. They allow to easily capture findings and their context. Correctly applied, this technique could improve feedback, while reducing customer effort. This paper collects our experiences from comparing acceptance testing with and without feedback systems in a distributed project. Our results indicate that this technique can improve acceptance testing – if certain requirements are met. We identify key requirements feedback systems should meet to support acceptance testing.

*Keywords*-distributed software development; requirements engineering; acceptance testing

## I. INTRODUCTION

Distributed software development projects are becoming the normal case nowadays [1]. This trend can be traced back to three major causes: economical, organizational, and strategic reasons [2]. However, distributed software development entails several challenges. One of these challenges is conducting acceptance tests [3]. Here, customers systematically use the system to determine if it meets all specified requirements[4], [5].

In this paper we use the term acceptance testing to describe a concise process that is comparable to an inspection [6], [7]. Accordingly, the customer can be seen as the reviewer, the acceptance test agent as the moderator, and the developer as the author. An acceptance test is executed interactively with the customer, produces a list of findings, and a final acceptance decision. Usually, acceptance tests can only be attended by few representative members of the project team, as these sessions are mostly carried out at a customer location. This introduces the challenge of sharing the customer's feedback among the team. In distributed projects this is even more challenging as the developers do not share a common context that helps them understand the feedback. Customer feedback is only indirectly transferred to the development team. Important information may be lost [8] or insufficiently documented [9]. Usually it is difficult and time consuming to describe a finding in sufficient detail. The fact that most findings are only useful and understandable if enough context information is given leads to our *problem statement*: If customer feedback or context information is lost during acceptance testing and the documentation of findings, the customer satisfaction is endangered.

One way to alleviate this problem is to integrate a dedicated feedback system into the system under construction to support and encourage the precise documentation of findings directly from the customer. However, the feature set of such feedback systems needs to be selected deliberately to not confuse or distract customers and to minimize the ambiguity of findings. To gather insight into requirements for feedback systems, we evaluated the application of a feedback system during acceptance testing in a distributed student project.

**Contribution:** In this paper we report our findings from a case study in which we used a feedback system for acceptance testing in a distributed software project.

- We share our **experiences** and findings from comparison with acceptance tests performed without a feedback system.
- Based on our experiences, we derive **requirements** a feedback system should fulfill when used to support acceptance testing.

In the following section II we give an overview of the related work in the area of acceptance testing and feedback systems. In Section III we specify our research questions and our research method. We present the empirical investigation of our study in Section IV. Finally, we discuss the requirements for feedback systems in Section V and conclude the paper in Section VI.

## II. RELATED WORK

Many *tools* offer mechanisms to gather information from users or to share information between them. The variety is broad, ranging from simple web forums [10] and web forms [11] to incident reporting tools [12] and integrated communication channels in software applications [13].

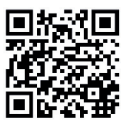

Stevens et al. [13] describe PaDU, an Eclipse plug-in for gathering information from end users. The tool allows users to report problems or suggestions within the application. They can create screenshots, annotate them and attach own sketches. The approach assumes that developers collect feedback and maintain a software system. Hartson et al. [12] describe a method for remote software evaluation in an external tool. They focus on reporting usability problems.

Çetin et al. [14] analyze how Human-Computer Interaction (HCI) experts can be involved in open source development. They identify essential requirements for HCI expert tools in distributed projects. They argue that these tools need to be used early in the development process.

Humayoun et al. [15] discuss the tool User Evaluation manager (UEMan) for the definition and deployment of experiments. Applied to requirements elicitation and user-centered design, it allows planning and automating evaluation experiments including code traceability. Due to tight integration of system and application, developers can see the results directly in their development environment.

Several approaches further aim at improving feedback processes. One aspect is the *context* in which a user can provide feedback. Jones et al. [16] and Krabbedijk et al. [17] propose workshops for multiple users. Their workflows help users to generate and communicate ideas for future system development. Another approach is to support feedback while a user uses a system and experiences drawbacks or has new ideas. Both, Seyff et al. [18] and Schneider [19] describe such approaches realized with mobile devices. This paper focuses on acceptance test session situations. This difference in the context of use leads to interesting new requirements as discussed in Section V.

Another aspect is the *technical environment* for the organization of user feedback. Castro-Herrera et al. [20] use forums for user feedback. They support the process by grouping ideas with data mining techniques and promoting forums with recommender systems. Lohmann et al. [21] leverage wikis and let users create and link requirements themselves. Our approach is not bound to a specific technical environment. Also, we do not support the organization and linking of requirements by users themselves. The idea is to keep the customers' burden during test sessions as low as possible.

Acceptance testing still involves sharp subjectivity and ad-hoc execution [22]. The customer uses the developed software application in order to determine if it meets the specified requirements [4], [5], [23]. Buede [24] discusses what should be tested and also includes noteworthy usability characteristics for this phase. Gibbs [25] describes checklists for the different roles involved in acceptance testing of outsourced projects.

In this paper we use an inspection as a comparable test procedure as described in [6], [7]. This definition allows to observe acceptance tests at the end or in the middle of a project, even to the point of test conduction after each completed user story in an agile project. Therefore, we try to discuss acceptance testing in a process agnostic way and only define the generic roles listed in Table I. Specific process models and roles can be mapped to these concepts.

### III. RESEARCH OBJECTIVE AND QUESTIONS

Our **research objective** is to apply a feedback system during acceptance testing in a distributed project and to evaluate its benefits and drawbacks. We do this by means of a case study based on the Goal Question Metric paradigm (GQM, [26]). Based on this research method, we derive the following **research questions** (RQ) for ourselves. For each research question we give the perspective from which we approach the question and the quality aspects, we consider relevant for the according perspective (c.f. [26]):

- RQ 1: Does the feedback system improve feedback? (perspective: *developer*; quality aspects: understandability, quantity, quality)
- RQ 2: Does the feedback system improve feedback? (perspective: *requirements owner*; quality aspects: quality, time needed)
- RQ 3: Does the feedback system lead to more customer satisfaction? (perspective: *requirements owner*; quality aspects: transparency, appropriateness)
- RQ 4: What are the key requirements for feedback systems in the context of acceptance testing?

We answer these research questions by performing the following steps:

- Integrate a *specific feedback system* into the software that is developed in a *specific distributed project*.
- Derive reasonable metrics to answer the research questions in our specific *study design* based on the GQM paradigm [26].
- Use observers (backed up by video), questionnaires, and interviews to capture metrics during acceptance test sessions (see Section IV).
- Derive requirements from the observations and discuss them with participants (see Section V).

For creating a baseline, we divide each acceptance test in two parts. One part, the *control part* is performed in a classic way: The customer performs acceptance tests supported by quality agents from the development team. The customer reports findings and the quality agents document these. The other part, the *test part*, is performed supported by the feedback system. The customer uses the feedback system to type in and submit findings.

### A. Specific Feedback System (FS)

To evaluate how a feedback system can affect acceptance testing in distributed projects, we chose a specific system

Table I
ROLES INVOLVED IN ACCEPTANCE TESTING.

| | Role | Description | # | Location | concrete allocation |
|---|---|---|---|---|---|
| 1. | Requirements Owner | Can determine to what extend a requirement is fulfilled | 2 | LUH | Academic staff |
| 2. | Decision Maker | Decides whether a new finding is created | | | |
| 3. | Developer | Knows existing solution and can estimate the impact of changes | 12 | RWTH | students |
| 4. | Acceptance Test Agent | Moderates the acceptance test session | 5 | LUH | students |

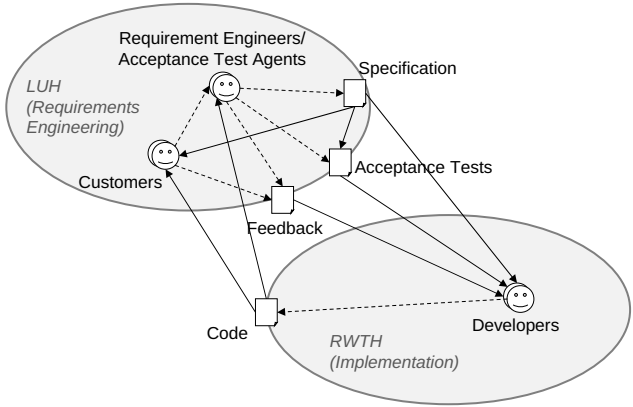

Figure 1. Situation and Context of Evaluation visualized in a FLOW Map (c.f. [29]).

[27]. For our research it is important that it is easy to integrate into the developed software and that it incorporates a useful set of features. We chose a tool that was developed at RWTH Aachen [28] because its features were a good starting point for our purposes. As the evaluation was mainly carried out in Hanover, we consider the additional threat to validity of using a self-made tool to be acceptable.

The FS supports different platforms by providing clients for applications on the Web, and in Eclipse and Android frameworks. It allows the user to draw on a web page, take a screenshot of the result, add comments, and submit this feedback to a ticket system in the back-end of the feedback system. Again, different ticket systems are supported. In our specific configuration we used the JavaScript client for Web pages and the Trac[1] ticket system.

### B. Specific Distributed Project

The study was conducted in a lab class distributed over four German universities (cf. [30]). The given task was to create a social network for distributed software projects. As our research questions only cover the customers', acceptance test agents', and developers' perspectives, we focus on the sites responsible for requirements analysis, acceptance testing, and implementation. Figure 1 shows the relevant parts of the project setup.

Following the waterfall process, requirements engineers started with requirements analysis and customer interviews. They wrote use cases and derived acceptance tests from use cases for the specification at the LUH site. Then, the developers at RWTH implemented the system. The code resided at the RWTH site. Afterwards, the customers and acceptance test agents at LUH conducted acceptance tests and documented the customers' feedback as Trac tickets and via the feedback system.

[1]trac.edgewall.org

The customers were located at LUH, represented by two individuals, *customer A* and *customer B*. *Customer A*, also referenced as the main customer, had the business case for the developed system and wanted to use it for a research project. The secondary customer, *customer B*, was also a contact person for the requirements engineers. Having knowledge about *customer A*'s business case, she acted as a customer proxy and second stakeholder.

Table I shows the concrete allocation inside the project. Both customers were represented by academic staff with considerable project experience from several industry projects. The requirements engineers and acceptance test agents held a Bachelor's degree in computer science at the time of the case study.

### C. Study Design

Our GQM goal for this study was to improve acceptance tests in distributed software projects by using integrated feedback systems (c.f. [26]). We defined *improvement* based on the perspectives of customers and developers.

To reduce learning effects we partitioned the test case set into four groups (ATG1 - ATG4), Table II. Each test session included two different runs, one with utilization of the feedback system (Mode FS) and one without FS as a control run (Mode CR). In both runs, test agents guided the customer through the acceptance tests. In each run, a customer executed two of the acceptance test case groups (ATGs), so that customers were not confronted with the same test case twice. Later, the results of the different runs are compared to each other to identify effects of adding the feedback system.

Table II
CROSS DESIGN OF OUR STUDY

| Acceptance test case group | ATG1 | ATG2 | ATG3 | ATG4 |
|---|---|---|---|---|
| Mode FS | A D1 | B D1 | A D2 | B D2 |
| Mode CR | B D2 | A D2 | B D1 | A D1 |

TABLE III
MEASUREMENT INSTRUMENTS

| |
|---|
| Minutes with observations (*Minutes-Obs*) were taken by two of the paper's authors during the acceptance test sessions. They document all mentioned findings (documented and not documented) and the times of announcement and documentation of a finding. |
| A questionnaire for developers (*Quest-DEV*) was filled out after all findings had been entered into the Trac system. The developers had access to the Trac system and answered questions regarding the findings. Per session six developers participated. |
| A questionnaire for the requirements owner (*Quest-RO*) was filled out after all findings had been entered into the Trac system. The questions cover the findings, track tickets, and the acceptance test itself. Per test session one requirements owner participated. |
| A questionnaire for testing agents (*Quest-TA*) was filled out after the acceptance test. Questions about the acceptance of the different test cases were asked. Per session two to three testing agents participated. |
| Trac tickets (*Trac-analysis*) were analyzed after the whole experiment. |
| Information flows (*FLOW-analysis*) were analyzed for the different runs. |

The session started with the *control run* (CR) that was performed without the feedback system. Here, the customers verbally reported findings and one test agent protocolled the mentioned findings with pen and paper. Later, a test agent transcribed the findings from the protocol into Trac tickets.

The second half of the test was performed with the feedback system (FS). The customer talked to the test agents, but this time documented findings by herself. For a finding, the customer filled out a form and optionally used built-in functions to make a screenshot and draw on it. Submitting the form with the FS, automatically created a Trac ticket.

In addition, we divided the developers into two groups (D1 and D2) of three developers. Each group received findings an ATG exactly once, switching between findings from the two customers and the two runs. We assume that test cases are more likely comparable than customers.

In order to improve the internal validity of our study, we tried to reduce the impact a specific customer, developer, or acceptance test agent could have on the results. We created different groups in order to exclude lerning effects. (The feedback system must be able to support unexperienced users.) Table III lists the measurement instruments we used for our study.

## IV. EMPIRICAL INVESTIGATION

In this section we describe the empirical investigation of the first three research questions. Often, an empirical observation leads to a requirement for feedback systems in acceptance tests (c.f. RQ 4). We highlight the requirements and give an overview of all requirements in Section V.

### A. Does the feedback system improve feedback from the developers' perspective? (RQ 1)

For a developer who did not attend the acceptance test, good documentation of the feedback is essential. We approached this research question with the aspects *understandability*, *quantity*, and *nature* of documented findings. Table IV displays the aspects and their operationalization.

The developers must understand all findings and their importance because otherwise they do not know what they have to do. A higher quantity of documented feedback is desirable to help the developers to improve the software. Further, critical findings have a higher impact on the final acceptance and require more attention. We define a finding as critical if it hinders acceptance of a test case.

*1) Measurement:* We used a questionnaire to assess the developers' perceived *understandability*. We have four possible combinations of understandable/not understandable findings with and without further questions. We computed the fractions of each group, once for findings documented with the feedback system (FS) and then for findings documented manually in the control run (CR). For the *quantity* of documented findings we counted and averaged the number of findings per test case within the Trac system. In a questionnaire requirements owners stated their perceived *criticality* for a finding and divided the findings into test failures and new defects. Then, we averaged the results.

*2) Results:* Figure 2 illustrates the perceived understandability of findings. Of the findings documented with the FS, 70% were considered understandable without any further questions and 23% understandable with further questions. This means that 93% of the findings were stated generally understandable. For manually documented findings, 79% were considered understandable without further questions – more than for findings documented with the FS. However, the fraction of all understandable findings (with and without questions) is only 84%.

The numbers of documented findings, divided by test sessions, are presented in Figure 3. Generally, with the FS, 6% more findings were documented per test case. However, the numbers of documented findings strongly deviate for the different sessions. In *session 2A*, 67% more findings per test case were documented with the FS. In *session 1B* in contrast, 33% more findings were documented during the control run.

Table V displays the findings' nature. 58% of the findings from the FS were considered critical. For manual documen-tation, the critical fraction is slightly higher (67%). At the same time, 60% of the findings from the FS and only 42% of manually documented findings reveal new defects.

TABLE IV
OPERATIONALIZATION OF RQ 1

| *Quality Aspect* | *Method* |
|---|---|
| Understandability of findings documented in Trac | |
|     Is the finding understandable? | Quest-DEV |
|     Are there further questions regarding the finding? | Quest-DEV |
| Quantity of findings documented in trac | |
|     Number of findings documented in Trac | Trac-analysis |
| Nature of findings documented in trac | |
|     Is the documented finding critical? (yes/no) | Quest-RO |
|     Type of finding (failed test / new defect) | Trac-analysis |

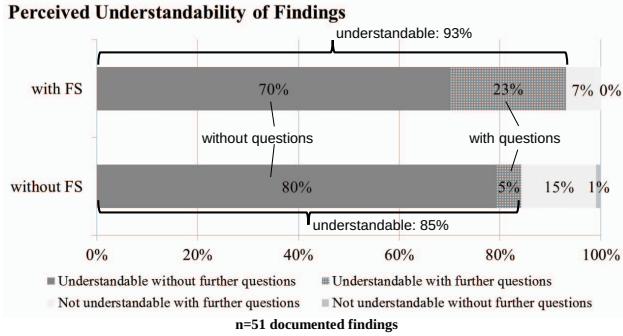

Figure 2. Perceived Understandability of Findings

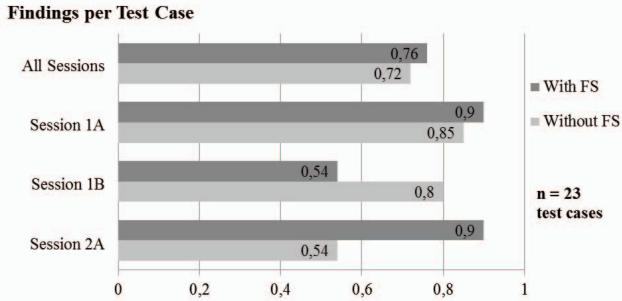

Figure 3. Findings per Test Case

*3) Discussion:* With the feedback system, acceptance tests are more user driven and more user interface oriented. This can have positive effects but also requires attention.

Findings from the FS were generally more understandable than those documented manually. However, at the same time, developers had more questions to FS-documented findings. We assume that questions to non-understandable findings aim at better understanding the findings whereas questions to understood findings aim at clarifying further details. Accordingly, the GUI centered documentation might have encouraged the developers to ask more clarifying questions. We do not have qualitative data on the question content though and therefore only can speculate.

The user interface orientation also influences distraction from the actually tested functionality. Acceptance tests focus on attesting main aspects of the software. Compared to manual documentation, the documentation with the feedback system however produced fewer critical findings and more newly spotted defects. This result emphasizes how important it is to support the requirements owner in focusing on the test cases. Nevertheless, the ability to reveal new misconceptions before deploying a new feature is valuable as well.

Table V
NATURE OF FINDINGS

| (n=51 documented findings) | With FS | Without FS |
|---|---|---|
| Percent of documented findings that are critical | 58% | 67% |
| Percent of documented findings that describe new defects | 60% | 42% |

Another problem was revealed by the high variability of the number of documented findings. Documentation with the FS is more user centered but at the same time depends more on the user and her ability to judge a product. The sessions 1A and 2A were both performed by *customer A*, session 1B by *customer B*. While *customer A* was the main customer and had produced the requirements during analysis, *customer B* was only a secondary customer for the product. During the experiment, we observed that *customer B* was undecided whether her findings were bugs or *customer A*'s desired features. As a result, she mostly documented her findings only when the moderating test agent told her to. As described by Rumpe et al. [31], the customer's willingness and ability have a major impact on the project in general.

*4) Derived Requirements:* We are confident that good tool support could make the testing process better. Such a tool should fulfill the following requirements:

- **Req-1:** Allow to reference a test case when documenting a finding.
- **Req-2:** Provide information about the test case currently executed.
- **Req-3:** Allow to decide upon the test case acceptance based on a set of collected findings.
- **Req-4:** Encourage users to explicitly assign criticality to a finding.

*B. Does the feedback system improve feedback from the customer's perspective? (RQ 2)*

When participating in acceptance tests, the customer needs an efficient way to make an acceptance decision and communicate findings. Every involved person should know which test cases are accepted or not, as in *certainty regarding acceptance*. The aspects *recall*, *directness*, *perceived quality of feedback*, and *duration* address the efficient creation and communication of findings. Table VI shows the relevant aspects for this research question.

A high recall of feedback means that many of the mentioned findings are documented and can be retrieved. Mentioned feedback that is not documented might not reach the distant developers. Directness addresses a similar problem. Indirect feedback that has been processed by different persons might include misunderstandings and loss of information.

*1) Measurement:* We used questionnaires to ask all participants (divided by roles) about the acceptance status of the test cases and their certainty regarding these anwers. We identified for which fraction of test cases *all* participants had the same understanding. Another questionnaire assessed the customers' perceived quality of the resulting findings in trac.

Table VI
OPERATIONALIZATION OF RQ 2

| | |
|---|---|
| Certainty regarding acceptance<br>   Is the test case accepted?<br>   How sure are you regarding the acceptance status of this test case? | Quest_DEV,<br>Quest_RO,<br>Quest_TA |
| Recall of feedback<br>   Ratio of documented findings to all findings (documented and not documented) | Minutes_Obs,<br>Trac_Analysis |
| Directness of creating findings<br>   How many stations exist where information can be lost or falsified | FLOW_analysis |
| Duration of creating a finding<br>   Time to create a ticket for a finding<br>   Time to document a finding | Minutes_Obs,<br>Trac_Analysis |
| Perceived quality of findings<br>   How well are the findings captured in the documentation?<br>   Is the criticality of the finding captured correctly? | Quest_RO |

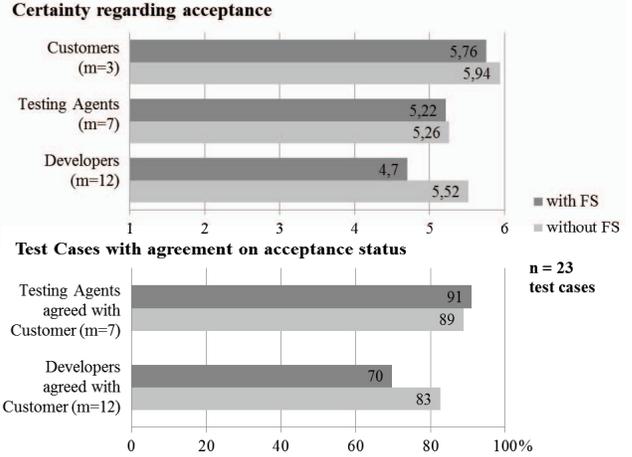

Figure 4. Certainty of Acceptance

Table VII
RECALL OF FEEDBACK

| Average percentage of mentioned findings that were documented as a ticket. n=71 mentioned findings | With FS | Without FS |
|---|---|---|
| acceptance test session 1A | 82% | 85% |
| acceptance test session 1B | 41% | 53% |
| acceptance test session 2A | 100% | 100% |

We took minutes of the test sessions and documented all verbally mentioned findings and the timestamps of their documentation (start and end of documentation in the feedback system). We compared the verbally stated findings with the findings actually documented in trac tickets. Further, we retrieved the ticket creation timestamps from trac.

To visualize the directness of feedback we created an information flow diagram (FLOW, [32]) of the two acceptance test situations. A connection illustrates an information flow between two points. At every new point, information can get lost or falsified.

*2) Results:* Figure 4 illustrates the certainty aspect. The stated certainty (first diagram) is always higher for the control run. Most values are close to 5.5 (on a scale from 1 (very uncertain) to 6 (very certain)) with customers having the highest certainty. While testing agents' certainty was slightly higher for the control run, their opinions actually conformed to the customer's opinion slightly less often during this run (second diagram). The absent developers show the highest differences, indicating better understanding of the acceptance status for the control run.

Recall of feedback is illustrated in Table VII. At two sessions the customers mentioned findings that were not persisted. The recall of findings for session 1A is 82% with FS and 85% in the control run. For session 1B the recall is 41% with the feedback system and 53% without.

The two acceptance test modes are illustrated in the FLOW diagram in Figure 5. With the FS the requirements owner directly creates the persistent finding. In contrast, in the control run the requirements owner tells a finding to the test agent who notes it down in a report. Only then findings are persisted from the report's notes.

A related aspect is the time to create findings in Table VIII. The time for ticket creation (:= time difference between creating an empty findings-form and saving the completed finding) was two minutes faster with FS. The total time to document a finding (:= time difference between the mention of a finding and the saving of the completed finding) was on average 51 seconds with FS and more than eight hours for manual recording and a (potentially delayed) transcription by the test agent.

The quality of the tickets as perceived by the requirements owners is illustrated in Figure 6. Regarding the question how well the message was represented by the tickets from the FS, the customers on average answered 3.1 on a scale from 1

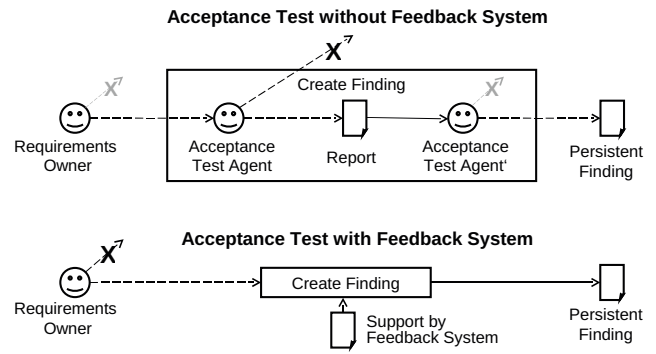

Figure 5. Directness of Creating Findings

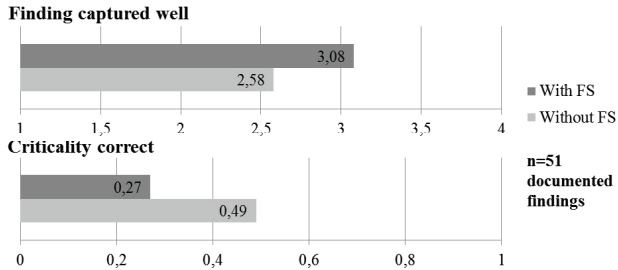

Figure 6. Perceived Quality of Findings

(very badly) to 4 (very well). The criticality was represented correctly for only 27% of the findings. For the tickets created manually in the control run, the average quality vote was 2.6. About 50% of the findings had correct criticality.

*3) Discussion:* With the feedback system tickets are persisted faster and created more directly by the customer herself without any intermediaries. Especially directness is important for communication in distributed environments where information can easily be lost or falsified. Fittingly, customers rated the quality of their directly written tickets higher than of the other tickets. However, the results reveal that recall of feedback and certainty regarding test case acceptance were better for the control runs without FS.

For manual documentation in the control run, the test agent had more freedom to include helpful information into tickets. The test agent grouped all findings by test case and included the acceptance state of each test case into the trac tickets. We assume that this improved the understanding of the acceptance status and the participants' certainty. In contrast, the FS form did not encourage the customer to document a finding's criticality, its test case, or the overall acceptance status of test cases. The criticality was even set to the misleading value *critical* for every ticket and therefore was wrong for 73% of the tickets (see Figure 6).

According to the recall of findings not all comments are persisted - interestingly, even when the customer documents findings by herself. This suggests a filtering process that separates true findings from simple comments, as illustrated by the crossed-out connectors in the FLOW diagram in Figure 5. With the feedback system the customer directly decides which comments should be persisted. Otherwise this decision is in the hands of the testing agent.

In session 1B, relatively few findings were recorded at all. We believe this happened because *customer B* was not a requirements owner for the product, like discussed in *RQ 1*. Due to uncertainty about the requirements, *customer B*

Table VIII
DURATIONS OF CREATION AND DOCUMENTATION OF FINDINGS

| (n=51 documented findings) | With FS | Without FS |
|---|---|---|
| Avg Time to Create Ticket for Finding | 0:37 mins | 2:37 mins |
| Avg Time to Document Finding | 0:51 mins | 520:26 mins |

Table IX
OPERATIONALIZATION OF RQ 3

| Seriousness | |
|---|---|
| Did you feel taken seriously throughout the acceptance test? | Quest_RO |
| How sure are you that all mentioned findings have been protocolled? | |
| Perceived bindingness of overall acceptance test | |
| How binding do you assess the acceptance test? | Quest_RO |

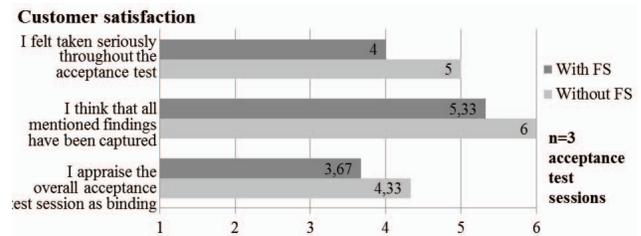

Figure 7. Customer Satisfaction

mainly made simple comments instead of declaring them as findings to be persisted.

*4) Derived Requirements:* Besides the finding itself further information is required to make findings more understandable and provide a better overview. The feedback system enriches findings with screenshots, but also needs to support adding criticality, acceptance information, and linked test cases, as stated in the following list.

- **Req-2:** Provide information about the test case currently executed.
- **Req-5:** Allow the assignment of a criticality value to a finding.
- **Req-6:** Display the acceptance status of a test case.
- **Req-7:** Display the overall acceptance of the test.

*C. Does the Feedback System Lead to More Customer Satisfaction? (RQ 3)*

Table IX shows the relevant aspects for this research question. Acceptance tests are an important and official part in the development lifecycle that must have a professional appearance. A customer should always feel that she and her feedback are *taken seriously*. The acceptance decision and the findings must have a high *bindingness*, indicating that the revealed misconceptions will actually be changed.

*1) Measurement:* In a questionnaire right after the test sessions the customers stated their subjective overall opinion about the two session parts – the control run without FS and the part with FS.

*2) Results:* The results for this questionnaire are listed in Figure 7. The average results are all between 4 and 6 on a scale from 1 (strong disagree) to 6 (strong agree). For all three questions the customers indicated a higher satisfaction for the control run.

*3) Discussion:* In a subsequent interview with the customers we found out that the contact with the test agent

gave them the feeling of being heard and taken seriously. When they only typed-in their findings into a system and pressed a button, they missed this contact. Especially the lack of a concluding message about what will happen to the feedback was perceived as unsecuring.

*4) Derived Requirements:* The system must indicate the act of testing and provide the current acceptance status. The submission must be emphasized. Feedback regarding the processing of a finding is essential to articulate its bindingness. We derive the following concrete requirements:

- **Req-6:** Display the acceptance status of a test case.
- **Req-7:** Display the overall acceptance of the test.
- **Req-8:** Pursue high transparency of the processing status of findings throughout their life cycle.
- **Req-9:** Provide an overview of the reported findings.
- **Req-10:** Provide feedback on what happens to a submitted finding.
- **Req-11:** Display explicitly to the customer that a test is being executed.

*D. Threats to Validity*

In this section we discuss threats to validity. By this, we want to support the correct interpretation of our results. We would classify our empirical investigation as *applied research*. According to Wohlin et al. [33], we set the highest priority on achieving good internal validity. Therefore, we explicitly addressed *internal validity* when we planned the experiment and describe the related activities with the study design in Section III-C.

*1) External Validity:* In our study, developers and test agents were graduate students. Both customers were academic staff instructed to always play the customer role when communicating with the students. The main customer had a real business case for the system under construction. He intended to use the system in his research projects and wanted a working system that fulfilled as many requirements as possible. The secondary customer had only limited decision power. The students had good background knowledge in software engineering but limited experience. This setting is typically encountered in an industrial setting. The main difference is the lack of possible monetary loss in our setting. Still, the students had an interest in passing the class.

*2) Construct Validity:* Our students can be seen as tomorrow's IT specialists. They grew up with feedback systems integrated in social software. They have domain knowledge and a general understanding on how web based social networks are used. The main customer was interested in a positive outcome. We prevented co-location with actual distribution between customers and developers. Our setup also prohibited additional information flows between customers and developers.

*3) Conclusion Validity:* In our evaluation, we did not try to achieve a statistical significance. We cannot guarantee that replicating our experiment will lead to similar results.

The questionnaires after the test sessions hold the risk that only subjective impressions are mentioned, but this holds for both modes. Only findings with a high criticality will be kept in mind by the customer (Quest_RO). The questionnaire of the developers (Quest_DEV) has the same condition for both modes, because the findings were presented in the same way to the developers.

Our results show that customers strongly influence the outcome. In our study we had two customers that represent two relevant customer types: the main customer was both requirements owner and decision maker, whereas the secondary customer was only decision maker. We think that both customer types are realistic in software projects. In addition, having two customers reduced our threats to internal and external validity.

## V. REQUIREMENTS FOR FEEDBACK SYSTEMS

Our results suggest that feedback systems can positively impact distributed acceptance tests. However, distributed acceptance test situations imply new requirements. In this section, we present the requirements derived from our case study in more detail and give guidelines on how to overcome the problems. We group the requirements by requirement types and discuss them together. Table X presents the mapping between requirement and requirement type.

*1) Background information about findings:* To raise a finding's understandability, its context should be provided. We differentiate between product context and execution context. Product context collects information of the tested product, like application and web browser version, screenshot, or finding description. Execution context describes the situation that led to a finding, like the test case, the executing person, the number and nature of identified findings.

Feedback systems should support generating context without burdening the users to enter much additional information. For example, if the customer chooses the current test case from a list before executing it, the feedback system can automatically link all findings to that test case. Further, customers should be able to attach annotated screenshots as known from other HCI studies [14] and end user feedback tools [11]. Findings should include a reference to the decision maker. Knowing who created a finding helps developers to better understand it and contact the creator in case of questions.

*2) Criticality:* Having a finding's criticality assigned by the customer helps to understand its importance. Customers should be encouraged to specify the criticality for each finding. A specific criticality field within a finding's form could advert to this necessity.

*3) Reference to test case:* The test case is part of a finding's context and increases understandability. Additionally, it should be presented to the customer by the feedback system. Seeing the current test case helps the customer understand the tested requirements and focus on what should be assessed. Later, links from a test case to its findings help the customer assess the test case's acceptance status, as described in the following section.

TABLE X
REQUIREMENTS MAPPED TO REQUIREMENT TYPES

| Req-No | Description | Type |
| --- | --- | --- |
| **Req-1** | Allow to reference a test case when documenting a finding | Reference |
| **Req-2** | Provide information about the test case currently executed | Reference, Background |
| **Req-3** | Allow to decide upon acceptance based on a set of collected findings | Status |
| **Req-4** | Encourage users to explicitly assign criticality to a finding | Criticality |
| **Req-5** | Allow the assignment of a criticality value to a finding | Criticality |
| **Req-6** | Display the acceptance status of a test case | Status |
| **Req-7** | Display the overall acceptance of the test | Status |
| **Req-8** | Pursue high transparency of the processing status of findings throughout their life cycle | Transparency |
| **Req-9** | Provide an overview of the reported findings | Status, Transparency |
| **Req-10** | Provide feedback on what happens to a submitted finding | Transparency |
| **Req-11** | Display explicitly to the customer that a test is being executed | Status, Background, Reference |

*4) Acceptance status of current test case:* Throughout the test, the acceptance status should be visible. We suggest a summary of the whole test and a separate status for each test case. Seeing all findings of one test case helps the customer determine its acceptance status. The customer might reject a test case due to critical findings, but also if there are too many minor findings. The feedback system should support to explicitly set a test case's acceptance status.

Also, a system should support the customer to determine the overall acceptance status. The feedback system could offer a *shopping cart* for findings. The cart summarizes all reported findings and allows to edit them and their criticality. To finish the test the customer would have to explicitly *checkout*, confirm all findings, and specify an overall acceptance statement.

*5) High transparency of acceptance test:* Transparency can increase a customer's feeling of being taken seriously. The customer should be able to follow her findings' life cycle and see if findings are fixed or dropped. Laurent and Cleland-Huang [10] state that users are interested in exactly this process of how findings are handled. In addition, a confirmation of the receipt of findings is very important, especially when the customers interact with a system instead of a person.

## VI. CONCLUSION

The increasing number of distributed software projects raises the communication need between project participants. We evaluated the application of an integrated feedback system in such a project. We focused on how feedback systems can support acceptance testing, especially by tackling the problem of lost feedback or context information. We covered the perspectives of different roles involved in this test. We found encouraging and also surprising results. The results point to the existence of several difficulties in acceptance testing. Especially the discussion of the results leads to interesting insights. Based on the experiences we identified 11 requirements for feedback systems. We grouped them into five different types and discussed possible solutions and concepts for feedback systems.

Our study shows that there is potential for using integrated feedback systems to support acceptance testing in distributed projects. Based on our insights, we plan to implement the new requirements in our feedback system. Then, another empirical investigation could lead to new valuable insights. Future research should focus on reducing the vulnerability of the process. The experiences and requirements presented in this paper are a good starting point for such efforts.

ACKNOWLEDGMENT

We thank academic staff from LUH, RWTH, TUC, and TUM (c.f. [30]) for organizing and their Students for participating in this distributed student's project and for providing clarifications when needed.

This work was partially funded by a Research Grant from the State of Lower Saxony, Germany.